\documentclass[twocolumn,aps,showpacs]{revtex4}
\usepackage{graphicx}%
\newcommand{\ba}{\begin{eqnarray}}
\newcommand{\ea}{\end{eqnarray}}
\begin{document}
\title{Ground state energy fluctuations in the Nuclear Shell Model}
\author{V\'\i ctor Vel\'azquez}
\affiliation{Dpto.de F\'\i sica, Facultad de Ciencias, Universidad Nacional Aut\'onoma de M\'exico, \\
Apartado Postal 70-542, 04510 M\'exico, D.F., M\'exico}
\author{Jorge G. Hirsch, Alejandro Frank, and Jos\'e Barea}
\affiliation{Instituto de Ciencias Nucleares, 
Universidad Nacional Aut\'onoma de M\'exico, \\
Apartado Postal 70-543, 04510 M\'exico, D.F., M\'exico}
\author{ Andr\'es P. Zuker}
\affiliation{IReS, B\^at27-CNRS/Universit\`e Louis Pasteur BP 28, F-67037 \\
Strasbourg Cedex 2, France\\}

\begin{abstract}
Statistical fluctuations of the nuclear ground state energies are estimated
using shell model calculations in which particles in the valence shells interact
through well defined forces, and are coupled to an upper shell governed by 
random 2-body interactions. Induced ground-state energy fluctuations are found to be one order of magnitude smaller than those previously associated with chaotic components, in close agreement with independent perturbative estimates based on the spreading widths of excited states.
\end{abstract}

\maketitle

The series of resonances found in the scattering of slow neutrons on heavy nuclei,
with typical excitation energies of 6 to 10 MeV, have energy-fluctuations described by
Wigner or Poisson distribution, depending on their quantum numbers \cite{Bro81}. 
These fluctuation properties of the quantal spectra \cite{Wig65} can be described in great detail using Random Matrix Theory \cite{Meh90}. The universality of the laws of level fluctuations
was postulated from the conjecture that classical chaotic systems have the same Wigner-type fluctuations properties as predicted by the Gaussian Orthogonal Ensamble (GOE) \cite{Boh84}, while classical integrable systems have Poisson-type distributions \cite{Boh85}. These results represent the strongest evidence of the presence of quantum chaos in nuclear systems \cite{Gut90,Sto99}. 
On the other hand, it has been convincingly shown that the power spectrum
(Fourier transform squared) of the fluctuations of a chaotic quantum energy spectra
is characterized by $1/f$ noise~\cite{Rel02}.

Extending these ideas to the nuclear ground state, and challenging preconceptions, it was recently postulated that the chaotic motion of the nucleons inside the nucleus induces chaotic fluctuations in the atomic masses \cite{Boh02}, of the same size of the deviations between experimental masses and those computed using the Finite Range Droplet Model (FRDM) \cite{Moll95},
of the order of 0.7 MeV.
This coexistence between order and chaos in the nuclear ground state was interpreted
as an inherent limit to the accuracy with which nuclear masses can be calculated~\cite{Abe02}.

The error distribution of the mass formulas of M\"oller et al. \cite{Moll95} have a conspicuous long range regularity \cite{Hir04,Hir04b} that manifests itself as a double peak in the distribution of mass differences \cite{Hir04}. This striking non-Gaussian distribution was found to be robust under a variety of criteria \cite{Hir04b}.
Detailed analysis of nuclear mass errors have been performed, using both the FRDM \cite{Moll95},
the shell-model inspired mass calculations of Duflo and Zuker (DZ)\cite{Duf94}, the Hartree - Fock- Bogoliubov mass calculations of Goriely et al. \cite{Gor01}, as well as the equations connecting the masses of a particular set of neighboring nuclei known as the Garvey-Kelson (GK)
relations~\cite{Gar66,Gar69}. 
The presence of strong correlations between mass errors in neighboring nuclei, calculated under a mean-field approach,
has been clearly exhibited, as well as the existence of a well defined chaotic signal in its 
power spectrum, when their correlations are analyzed as time series \cite{Hir04a,Bar04}. 
A similar power spectrum analysis demonstrates that the inclusion of additional physical contributions
in mass calculations, through many-body interactions or local information,
removes the chaotic signal in the discrepancies between calculated and measured masses \cite{Hir04a,Bar04}. It has also been shown that the average mass error is smaller that 100 keV. It represents an upper bound on the influence of the presence of chaos, which could preclude precise deterministic predictions of nuclear masses.

Another relevant question is whether nuclear masses far from stability
can be predicted with a similar accuracy.
Large-scale shell model calculations shed some light on this important issue~\cite{Hon02}.
Their predictive power seems robust against long-distance extrapolations.
Masses of 67 light nuclei in the $fp$ shell, many of them unstable,
were calculated with an average error of 215~keV~\cite{Cau99}.
Errors come mostly from isospin violation, and do not increase far from stability.
Large-scale shell model calculations for the $N=126$ Po-Pu isotones~\cite{Cau03}
find errors in the binding energies smaller than 50~keV
along the whole chain, with no increase for unstable nuclei,
and ``imply a high predictive power for ground-state binding energies
beyond the experimentally known nuclei.''

Knowing that the upper bound for the average error in both the calculability and the predictability of nuclear masses is smaller than 100 keV, it is still relevant to gauge the influence on the ground state of the``chaotic layers" associated with quantum chaos in nuclei. 

In order to understand the nature of the chaotic errors, in \cite{Vel03} a systematic study of fluctuations in nuclear masses was carried out using the shell model restricted to a single shell for protons or neutrons. Realistic Hamiltonians combined with a small random component were employed to study the fluctuations of the nuclear masses, and its dependence on the size of the Hilbert
space and on the width of the two-body interaction.
Ground state energy fluctuations were calculated for 39 nuclei in the {\it sd} and {\it pf} 
shells, using a combination of realistic and random 2-body interactions. Fluctuation widths were found to follow an approximate  $A^{-1/3}$ pattern \cite{Vel03}, similar to the one found using the periodic orbit theory \cite{Boh02}. Although these results exhibit the close relationship between the A-dependence of the 2-body interactions, the many body effects and the fluctuations in the ground state energy, both the origin and the size of these fluctuations were introduced in an artificial way.

The statistical fluctuations of the ground state energy induced by the stochastic behavior of levels in the vicinity of neutron threshold were investigated in \cite{Mol04} using both perturbation theory and supersymmetry. By employing the spreading width to estimate the mixing of the ground state with the states in the shell above, a relatively flat error (for A$>$50) close
to 100 keV was found. A more detailed calculation using second order perturbation theory and random 2-body interactions shows that the fluctuations decrease with increasing mass number, with
a functional form \cite{Abe05}
\begin{equation}
\sigma_{pert} = 19 \, A^{-4/3}\, \hbox{MeV}  . \label{s-pert}
\end{equation}

In the present letter we report the analysis of ground state energy fluctuations induced by random 2-body interactions, through large scale shell model calculations. The valence space includes two major shells for protons and two for neutrons. Single-particle energies are fixed, as well as both the residual interactions between particles in the lower shells and the interactions inducing mixing between shells. These forces are taken from the best available parameterizations in the literature. Two body interactions in the upper shell are taken randomly from a Gaussian distribution, whose width is determined from known interactions. Random matrices in the upper shell represent the presence of chaos in neutron resonances \cite{Mol04,Abe05}. 

\begin{figure}[h!]  
  \begin{center}
    \includegraphics[width=8.0cm]{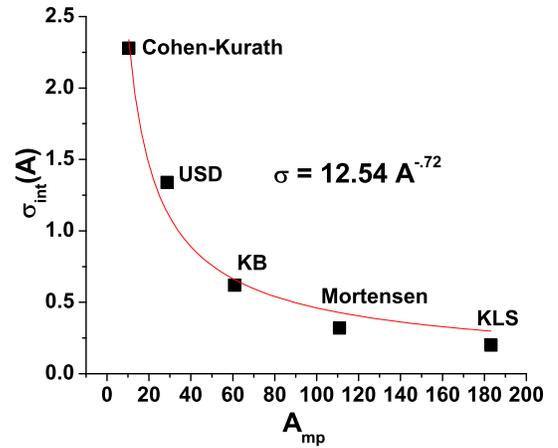}
  \end{center}
\caption{Average interaction widths as function of the middle shell $A_{mp}$.}
\label{anchura-int}
\end{figure}

To determine the widths of the distributions of the 2-body residual interaction
a statistical analysis of realistic interactions employed in shell model calculations, restricted to a single shell, was performed. These interactions are Cohen-Kurath for the $p$ shell \cite{CK}, Brown-Wildenthal for the $sd$ shell \cite{BW}, the KB3 interaction for the $pf$ shell\cite{KB3}, the BonnC interaction for the $gds$ shell \cite{Bonn}, and the KLS interaction for the $hf$ shell \cite{KLS}. As most of these interactions are fixed for a given shell, with no dependence on the mass number, their average width was associated with a representative mass number $A_{mp}$ at the middle of a given shell. A plot of the interaction widths $\sigma_{int}$ against the mass number was obtained in this way, which is shown in Fig. \ref{anchura-int}.
A simple fit suggest the functional form
\begin{equation}
\sigma_{int}(A) = 12.54 \, A^{-0.72} \, \hbox{MeV}.
\end{equation}

To estimate the ground state energy fluctuations, shell model calculations were performed using
the code Antoine \cite{Antoine,APP}.
Single particle energies were assumed fixed, as well as the interactions between particles in the lower shells, where the realistic forces optimized for shell model calculations were employed. The coupling between the lower and upper shells was built from the KLS interaction \cite{KLS}. The interactions in the upper shells were assumed random, with values taken from a Gaussian distribution centered at zero, and a width $\sigma_{int}(A)$. The opening of the upper shells significantly extends the Hilbert space and would imply a renormalization of the realistic interactions used in the lower shell. This effect is important for detailed spectroscopic calculations, but does not influence the energy fluctuations, so we will not give further consideration to this problem here.

\begin{figure}[h!]  
  \begin{center}
    \includegraphics[width=8.0cm]{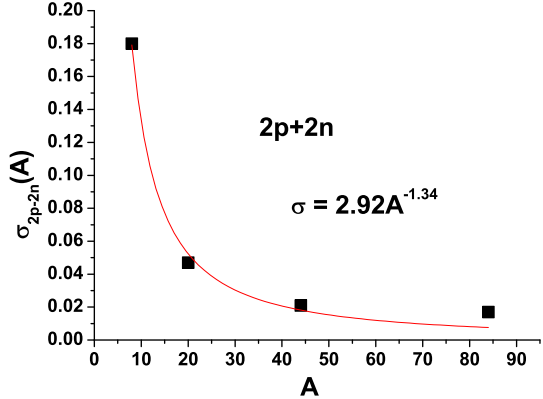}
  \end{center}
\caption{Ground state energy fluctuations for two protons plus two neutrons outside a closed core.}
\label{dospromasdosneu}
\end{figure}

Performing full shell model calculation in a space associated with many valence particles in four major shells (2 for protones and 2 for neutrons) is a formidable task. The size of the Hilbert space is so huge that in most cases they are impracticable even for the best available codes.
For this reason we performed the calculations for 2 protons and 2 neutrons outside closed shells
with N = Z = 2, 8, 20, 40.  The 4 representative nuclei studied were $^8$Be, $^{20}$Ne, $^{44}$Ti and $^{84}$Mo. The valence space included, respectively, the following combinations: $p-sd$, $sd-pf$, $pf-gds$, and $gds-hf$ shells. In the first two cases 1000 calculations were performed, 500 for the third, and 200 for the last case, varying in each calculation the numerical values of the interactions in the upper shells. As the energy fluctuations become smaller for the heavier nuclei, the smaller number of trials does not affect the quality of the results, shown in Fig. \ref{dospromasdosneu}.
The calculated ground state energy fluctuations in nuclei with two protons and two neutrons outside a closed core, $\sigma_{2p-2n}(A)$, induced by the use of random interactions in the upper shell, are shown as points in Fig. \ref{dospromasdosneu}. They can be parameterized with a smooth curve (full line) as
\begin{equation}
\sigma_{2p-2n}(A) = 2.92 \, A^{-1.34} \hbox{MeV}. 
\end{equation}
Notice that the A-dependence of the fluctuations of the 2-body interaction strengths behaves like $A^{-2/3}$, as seen in Fig. \ref{anchura-int}. Allowing these fluctuations to be present only in the upper shells, and to affect the ground-state energy through many-body effects in the diagonalization, leads to a $A^{-4/3}$ dependence in the case of the 2 protons and 2 neutrons outside a closed core. 

\begin{figure}[h!]  
  \begin{center}
    \includegraphics[width=8.0cm]{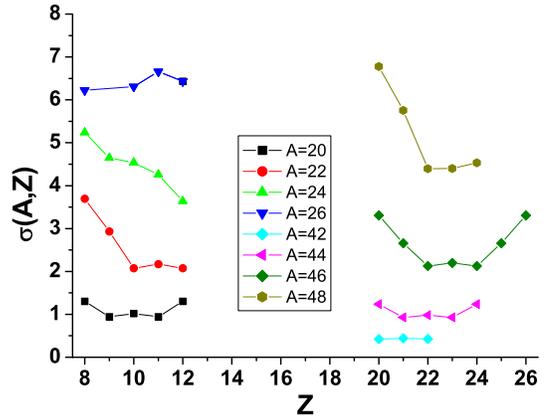}
  \end{center}
\caption{Mass widths in 20 nuclei in the sd- and fp-shells, relative to the 2p-2n width.}
\label{nto2}
\end{figure}

The final goal of this letter is to determine the size of the ground-state energy fluctuations for nuclei all along the periodic table, most of whom have many valence particles. For a majority of known isotopes, performing shell model calculations including two major shells for protons and two for neutrons is impossible. To circumvent this difficulty, we rely on the relationship between the g.s.-energy fluctuations of systems with 2 protons and 2 neutrons outside the core and those with $Z_v$ valence protons and $N_v$ valence neutrons obtained with statistical calculations performed with random interactions in one major shell \cite{Vel03}. This shell model study was restricted to a single shell for protons or neutrons, using a realistic Hamiltonian combined with a random component. We studied the ground state energy fluctuations of 39 nuclei in the {\it sd} y {\it pf} 
shells. It was shown that, for a wide range of mixing, the fluctuations grow linearly with the size of the random component. In this regime, the ratio between the fluctuations in nuclei with $Z_v$ valence protons and $N_v$ valence neutrons and those with 2 valence protons and 2 valence neutrons ranges between 1 and 7, where the largest ratio corresponds to mid-shell nuclei. The results for a 10\% random mixing within a major shell are described in detailed in Fig. 7 and Table II of Ref. \cite{Vel03}. They are summarized in Fig. \ref{nto2}. 
In what follows the conservative assumption of a factor of about seven between the mid-shell fluctuations and those found for 2p-2n calculations is employed. 

\begin{figure}[h!]  
  \begin{center}
    \includegraphics[width=8.0cm]{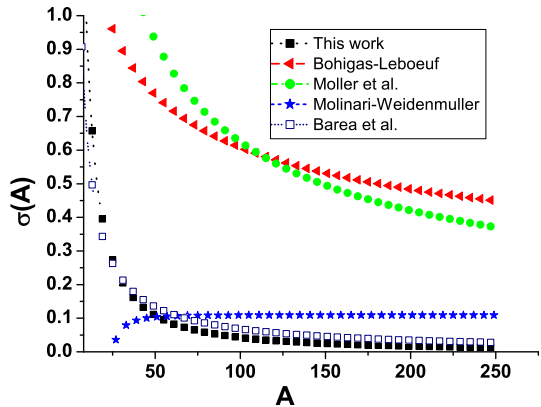}
  \end{center}
\caption{Comparison of theoretical mass errors in the FRDM and in GK calculations with different
estimates of ground-state energy fluctuations, color on line. }
\label{comparacion}
\end{figure}

By applying all the pervious results we conclude that the ground-state energy fluctuations induced by random interactions in the upper major shell are of the order
\begin{equation}
\sigma(A) \approx 20 \, A^{-1.34} \, \hbox{MeV}. 
\end{equation}
This estimate of the random ground-state energy fluctuations, shown with full squares in Fig. \ref{comparacion}, has some interesting features. They are better appreciated by comparison to the different results for the mass errors and fluctuations presented in Fig. \ref{comparacion}.
For a mass $A=100$ the induced chaotic fluctuations are of the order of 40 keV. Its mass dependence is very close to $A^{-4/3}$. The average r.m.s. mass error in the FRDM \cite{Moll95}, shown with full circles in Fig. \ref{comparacion},  were interpreted as a chaotic mass component in \cite{Boh02}, whose estimated errors are depicted with triangles.  They are both an order of magnitude larger than the present estimates, and decrease more slowly as a function of A. 
The curve with stars presents the perturbative calculation of Molinari-Weidenm\"uler \cite{Mol04}, with an odd dependence in the mass number as calculated in this reference. 
The fluctuations estimated in Ref. \cite{Abe05} using second order perturbation theory are so close to the present results that they are not displayed, since they would superimpose.

It is worth mentioning that local mass errors, deduced from the Garvey-Kelson relations, have the same average size \cite{Bar04}, and have a very similar mass dependence \cite{Bar04b}, shown with open squares, as the present estimate of chaotic energy fluctuations. A careful analysis would be necessary to disentangle the consequences of this coincidence.

Large shell model calculations including random interactions in the upper shells, which mimic the presence of quantum chaos associated with neutron resonances at energies between 6 to 10 MeV, have been performed. After gauging the average size of the interaction widths in different mass regions, diagonalizations were done for 2 protons and 2 neutrons in very large Hilbert spaces, including two major shells for protons and 2 for neutrons. A conservative scaling, based on random matrix studies in a single major shell, was employed to deduce an upper bound for the energy fluctuations at mid-shell. This estimate is consistent with the mass errors found in large shell model calculations along the N=126 line, and with local mass error estimated using the Garvey-Kelson relations, all being smaller than 100 keV. The mass dependence was fitted, and found to be close to $A^{-4/3}$. This agrees in both size and functional form with the fluctuations deduced independently from second order perturbation theory. Being an order of magnitude smaller than the mass errors in the FRDM, associated with a chaotic component in \cite{Boh02}, we conclude that this component cannot be related to {\em nuclear chaos} as defined by neutron resonances. The fact that this large chaotic component, whose existence was confirmed by a power spectrum analysis, disappears when additional 2-body forces or local mass information are employed, strongly suggest that it is 
nota physical phenomenon but rather a characteristic arising from the mean field approximation. Investigations along these lines are underway.

V.V. thanks the IReS for its hospitality.
This work was supported in part by Conacyt, M\'exico and DGAPA-UNAM.


\begin{thebibliography}{aa}
\bibitem{Bro81} T.A. Brody, J. Flores, J.B. French, P.A. Mello, A.Pandey, and S.S. M. Wong, Rev. Mod. Phys. {\bf 53} (1981) 385.
\bibitem{Wig65} E.P.~Wigner, {\it Statistical Theories of Spectra: Fluctuations}
(Academic, New York, 1965).
\bibitem{Meh90} M.L.~Mehta, {\it Random Matrices} (Academic Press, London, 1990).
\bibitem{Boh84} O.~Bohigas, M.J.~Gianonni, and C.~Schmit, Phys.\ Rev.\ Lett. {\bf 54}, 1 (1984).
\bibitem{Boh85} O.~Bohigas, R.U.~Haq, and A.~Pandey, 
Phys.\ Rev.\ Lett. {\bf 54}, 1645 (1985).
\bibitem{Gut90} M.C.~Gutzwiller, {\it Chaos in Classical and Quantum Mechanics}
(Springer, Berlin, 1990).
\bibitem{Sto99} H.-J.~St\"ochmann, {\it Quantum Chaos: An Introduction}
(Cambdrige University Press, Cambridge, 1999).
\bibitem{Boh02} O. Bohigas, P. Leboeuf, Phys. Rev. Lett. {\bf 88}, 92502 (2002).
\bibitem{Moll95} P. M\"oller, J.R. Nix, W.D. Myers, W.J. Swiatecki,
At. Data Nucl. Data Tables {\bf 59}, 185 (1995).
\bibitem{Rel02}  A. Rela\~no, J.M.G. G\'omez, R.A. Molina, J. Retamosa and E. Faleiro,
Phys. Rev. Lett. {\bf 89} (2002) 244102;
%
E.~Faleiro, J.M.G.~G\'omez, R.A.~Molina, L.~Mu\~noz, A.~Rela\~no,
and J.~Retamosa,
Phys. Rev. Lett. {\bf 93}, 244101 (2004).
\bibitem{Abe02} S. \AA berg, Nature {\bf 417}, 499 (2002).
\bibitem{Hir04} J.G.~Hirsch, A.~Frank, and V.~Vel\'azquez, 
Phys.\  Rev.\ C {\bf69}, 37304 (2004).
\bibitem{Hir04b} J.G.~Hirsch, V.~Vel\'azquez, and A.~Frank,
Rev.\ Mex.\ F\'\i s. {\bf50} Sup 2 (2004), 40.
\bibitem{Duf94} J. Duflo, Nucl. Phys. {\bf A 576}, 29 (1994); 
J. Duflo and A. P. Zuker, Phys. Rev. {\bf C 52}, R23 (1995).
\bibitem{Gor01} S. Goriely, F. Tondeur, and J.M. Pearson, Atom. Data Nucl. Data
Tables {\bf 77}, 311 (2001).
\bibitem{Gar66}
G.T.~Garvey and I.~Kelson, 
Phys.\ Rev.\ Lett. {\bf 16}, 197 (1966).
%
\bibitem{Gar69}
G.T.~Garvey, W.J.~Gerace, R.L.~Jaffe, I.~Talmi, and I.~Kelson, 
Rev.\ Mod.\ Phys. {\bf 41}, S1 (1969).
\bibitem{Hir04a}
J.G.~Hirsch, V.~Vel\'azquez, and A.~Frank, Phys.\ Lett.\ B {\bf595}, 231 (2004).
\bibitem{Bar04} J. Barea, A. Frank, J.G.~Hirsch,and P. van Isacker, Phys. Rev. Lett.
in press.
\bibitem{Hon02}
M.~Honma, T.~Otsuka, B.A.~Brown, and T.~Mizusaki,
Phys. Rev. C {\bf 65}, 061301(R) (2002).
%
\bibitem{Cau99} Etienne Caurier, G. Mart\'\i nez-Pinedo, F. Nowacki, A. Poves, J. Retamosa, and A.P. 
Zuker, Phys. Rev. {\bf C 59} (1999) 2033.
%
\bibitem{Cau03}
E.~Caurier, M.~Rejmund, and H.~Grawe,
Phys.\ Rev.\ C {\bf 67}, 54310 (2003). 
\bibitem{Vel03} V\'\i ctor Vel\'azquez, Jorge G. Hirsch, and Alejandro Frank, 
Rev. Mex. F\'\i s. {\bf 49} S. 4 (2003)34-38.
\bibitem{Mol04} A. Molinari, H.A. Weidenm\"uller, Phys. Lett. {\bf B 601} 
(2004) 119-124.
\bibitem{Abe05} Sven \AA berg, Phys. Lett. {\bf B} submitted.
%
\bibitem{CK} S. Cohen and D. Kurath, Nucl. Phys. {\bf 73}(1965)1; {\it ibid} Nucl. Phys. {\bf A 101} (1967) 1.  
\bibitem{BW} B. H. Wildenthal,  Prog. Part. Nucl. Phys. {\bf 11} (1984)5. 
\bibitem{KB3} T.T.S. Kuo and G.E. Brown, Nucl. Phys. {\bf A 114} (1968)235; A. Poves and  A.P. Zuker, Phys. Rep. {\bf 70} (1981) 235.
\bibitem{Bonn} M. Hjorth-Jensen, T.T.S. Kuo, and E. Osnes, Phys. Rep. {\bf 261} (1995)125; R. Machleidt, F. Sammarruca, Y. Song, Phys. Rev. {\bf C 53} (1996)1483.
\bibitem{KLS} S. Kahana, H.C. Lee, and C.K. Scott, Phys. Rep. {\bf 180} (1969) 956; Phys. Rep.{\bf 185} (1969) 1378.
%
\bibitem{Antoine} E. Caurier, shell model code ANTOINE, IReS, Strasbourg, 1989-2004.
\bibitem{APP} E. Caurier and Frederic Nowacki, Acta Physica Polonica, {\bf{B30}}, 3(1999) 705.
\bibitem{Zuk04} Andr\'es P. Zuker, Rafael Molina, Marianne Dufour, Rev. Mex. Fis. 
{\bf 50 S4} (2004) 117-121. 
\bibitem{Bar04b} J. Barea, private communication.
%
\end{thebibliography}
\end{document}